\documentclass[aps,prd,reprint,preprintnumbers,nofootinbib]{revtex4-1}
\pdfoutput=1

\usepackage{graphicx}
\usepackage{amsmath}
\usepackage{amssymb}

\usepackage[colorlinks]{hyperref}
\hypersetup{allcolors=[rgb]{0,0,1}}

\newcommand{\orcid}[1]{\href{https://orcid.org/#1}{#1}}
\newcommand{\dmsqeea}{\Delta\wh{m^2}_{ee}}
\newcommand{\dmsqee}{\Delta m^2_{ee}}
\newcommand{\eps}{\epsilon}
\DeclareMathOperator{\tr}{tr}
\newcommand{\wt}[1]{\widetilde{#1}}
\newcommand{\wh}[1]{\widehat{#1}}
\newcommand{\e}[1]{\times10^{#1}}

\usepackage{scalerel}
\newlength{\heightnu}
\AtBeginDocument{\settoheight{\heightnu}{$\nu$}}

\begin{document}

\title{The Effective \texorpdfstring{$\Delta m^2_{ee}$}{Delta m squared ee} in Matter}

\author{Peter B.~Denton}
\email{peterbd1@gmail.com}
\thanks{\orcid{0000-0002-5209-872X}}
\affiliation{Niels Bohr International Academy, University of Copenhagen, The Niels Bohr Institute, Blegdamsvej 17, DK-2100, Copenhagen, Denmark}

\author{Stephen J.~Parke}
\email{parke@fnal.gov}
\thanks{\orcid{0000-0003-2028-6782}}
\affiliation{Theoretical Physics Department, Fermi National Accelerator Laboratory,\\ P.~O.~Box 500, Batavia, IL 60510, USA}

\preprint{FERMILAB-PUB-18-286-T}

\date{August 28, 2018}

\begin{abstract}
In this paper we generalize the concept of an effective $\Delta m^2_{ee}$ for $\nu_e/\bar{\nu}_e$ disappearance experiments, which has been extensively used by the short baseline reactor experiments, to include the effects of propagation through matter for longer baseline $\nu_e/\bar{\nu}_e$ disappearance experiments. This generalization is a trivial, linear combination of the neutrino mass squared eigenvalues in matter and thus is not a simple extension of the usually vacuum expression,  although, as it must, it reduces to the correct expression in the vacuum limit. We also demonstrated that the effective $\Delta m^2_{ee}$ in matter is very useful conceptually and numerically for understanding the form of the neutrino mass squared eigenstates in matter and hence for calculating the matter oscillation probabilities.
Finally we analytically estimate the precision of this two-flavor approach and numerically verify that it is precise at the sub-percent level.
\end{abstract}

\maketitle

\section{Introduction}
Since the discovery that neutrinos oscillate \cite{Fukuda:1998mi,Ahmad:2002jz} tremendous progress has been made in understanding their properties.
The oscillation parameters are all either well-measured or will be with the advent of next generation experiments.
As the final parameters are measured, precision in the neutrino sector becomes more important than ever.

In vacuum, an effective two-flavor oscillation picture was presented in \cite{Nunokawa:2005nx} for calculating the $\nu_e \rightarrow \nu_e $ disappearance probability which introduced an effective $\Delta m^2$,
\begin{equation}
\Delta m^2_{ee}\equiv\cos^2\theta_{12}\Delta m^2_{31}+\sin^2\theta_{12}\Delta m^2_{32}\,,
\label{eq:defn0}
\end{equation}
which precisely and optimally determines the shape of the disappearance probability around the first oscillation minimum. 
That is,
even in the three favor framework, for $\nu_e$ disappearance in vacuum ($P_0$), the two-flavor approximation
\begin{align}
P_0(\nu_e\to\nu_e):
& \approx 1- \sin^22\theta_{13}      \sin^2 \Delta_{ee} , \label{eq:P0}  \\
{\rm where} \quad \Delta_{ee} & \equiv \dmsqee L/(4E) \,,  \nonumber
\end{align}
is an excellent approximation at least over the first oscillation.
$\Delta m^2_{ee}$ has been widely used by the short baseline reactor experiments, Daya Bay \cite{An:2016ses} and RENO \cite{RENO:2015ksa} in their shape analyses around the first oscillation minimum and will be precisely  measured to better than 1\% in the medium baseline JUNO \cite{An:2015jdp} experiment.

The matter generalization of the three-flavor $\nu_e$ disappearance probability in matter ($P_a$) can also be adequately approximated by a two-flavor disappearance oscillation probability in matter
\begin{align}
P_a(\nu_e\to\nu_e)
& \approx 1- \sin^22\theta_{13}     \left( \frac{\Delta m^2_{ee}}{\Delta \wh{m^2}_{ee}} \right)^2 \sin^2 \wh{\Delta}_{ee}\,, \label{eq:Pa2} \\
{\rm where} \quad \wh{\Delta}_{ee} & \equiv \dmsqeea L/(4E) \,,  \nonumber
\end{align}
and $\wh x$ denotes the exact matter version of a variable and is a function of the Wolfenstein matter potential \cite{Wolfenstein:1977ue}.
This new $\Delta\wh{m^2}_{ee}$ would be the dominant frequency, over the first few oscillations, for $\nu_e$ disappearance at a potential future neutrino factory \cite{Albright:2000xi} in the same way that $\Delta m^2_{ee}$ is for short baseline reactor experiments.
As we will find in section \ref{sec:defn},
\begin{align}
\Delta \wh{m^2}_{ee} & \equiv \wh{m^2}_3-(\wh{m^2}_1 +\wh{m^2}_2) \nonumber \\
 &-[m^2_3-(m^2_1 +m^2_2)]  + \Delta m^2_{ee}  \label{eq:ee-intro}
\end{align}
satisfies all of the necessary criteria to describe $\nu_e$ disappearance in matter in the approximate two-flavor picture of eq.~\ref{eq:Pa2} above and trivially reproduces eq.~\ref{eq:defn0} in vacuum.   

We will also discuss an alternate expression $\Delta\wh{m^2}_{EE}$ which numerically behaves quite similarly, but is somewhat less useful analytically.

The layout of this paper is as follows.
In section \ref{sec:defn} we define the matter version of $\Delta m^2_{ee}$ denoted $\Delta\wh{m^2}_{ee}$.
We review the connection between the three-flavor and two-flavor expressions in section \ref{sec:three to two} which naturally leads to a slightly different expression dubbed $\Delta\wh{m^2}_{EE}$.
In section \ref{sec:relation to DMP} we show how the natural definition of $\Delta\wh{m^2}_{ee}$ matches the expression given from a perturbative description of oscillation probabilities.
We analytically and numerically show that both expressions are very close in section \ref{sec:two expressions}.
We perform the numerical and analytical calculations to show the precision of this definition of $\Delta\wh{m^2}_{ee}$ compared with other definitions of $\Delta m^2_{ee}$ in matter in section \ref{sec:precision}.
Finally, we end with our conclusions in section \ref{sec:conclusions}, and some details are included in the appendices.

\section{Defining \texorpdfstring{$\Delta\wh{m^2}_{ee}$}{Delta m squared ee} in matter}
\label{sec:defn}
In this section we create a qualitative picture to derive the $\Delta\wh{m^2}_{ee}$ presented in the previous section.
We then verify that it passes the necessary consistency checks.

\begin{figure}
\centering
\includegraphics[width=\columnwidth]{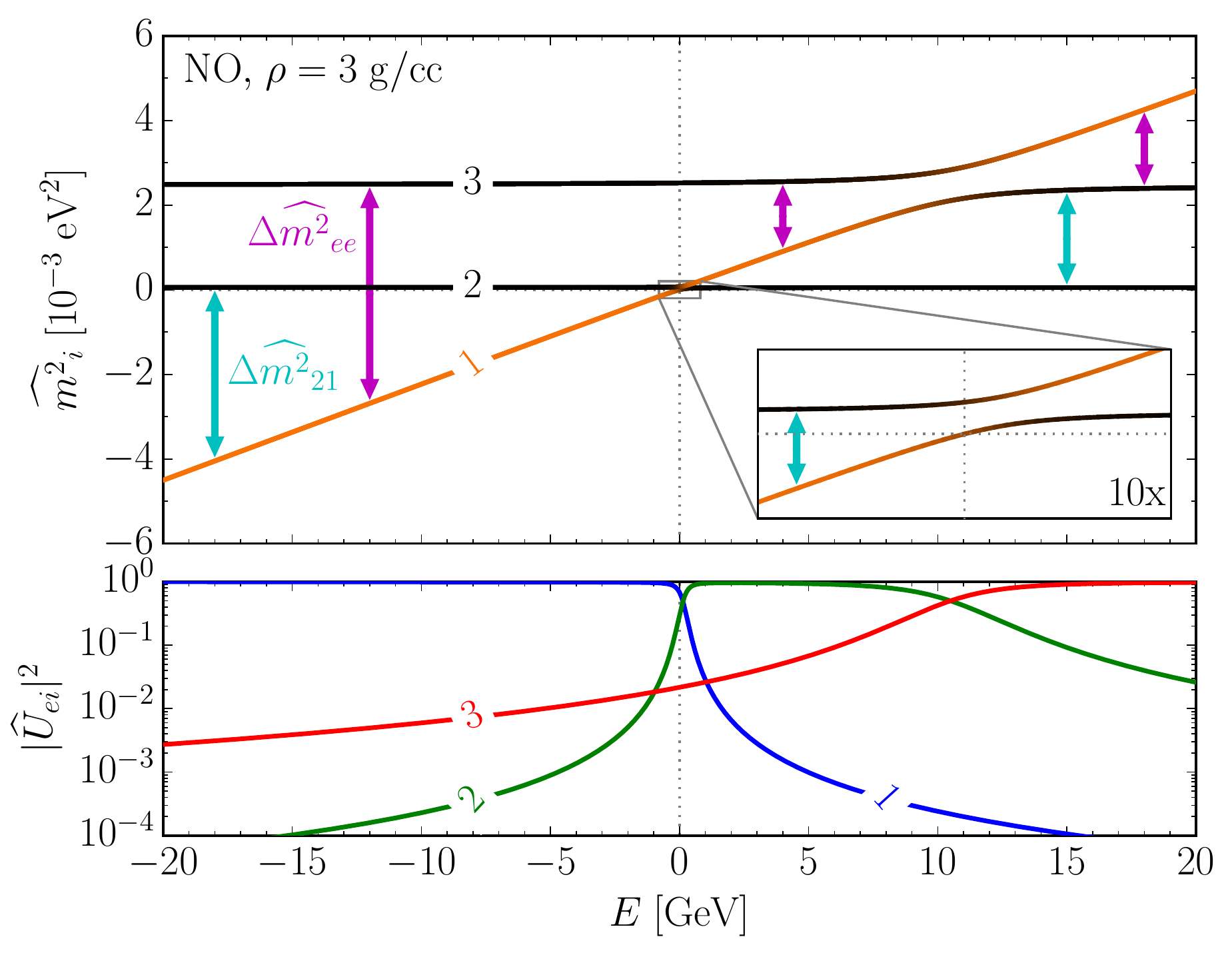}
\caption{Upper panel: the eigenvalues as a function of energy for $\rho=3$ g/cc and the NO.
Positive energies refer to neutrinos while negative energies refer to anti-neutrinos; $E=0$ refers to the vacuum.
The $\nu_e$ content of each eigenvalue is shaded in orange, while the $\nu_\mu$ and $\nu_\tau$ content is shaded in black.
The magenta (cyan) arrows indicate how $\dmsqeea$ ($\Delta\widehat{m^2}_{21}$) changes with energy.
Lower panel: the $\nu_e$ content of each mass eigenstate, $|\wh{U}_{ei}|^2$, as a function of neutrino energy.}
\label{fig:msqeds}
\end{figure}

Figure \ref{fig:msqeds} gives the neutrino mass squared eigenvalues in matter, $\wh{m^2}_i$, as a function of the neutrino energy as well as the value of their electron neutrino content, $|\wh{U}_{ei}|^2$.
Neutrinos (anti-neutrinos) are positive (negative) energy in this figure and vacuum corresponds to $E=0$. From the $\nu_e$ content, it is clear that  for energies greater than a few GeV that $\Delta \wh{m^2}_{32} $ will dominate the L/E dependence of $\nu_e$ disappearance and similarly $\Delta \wh{m^2}_{31} $ will dominate for energies less than negative, a few GeV, that is,
\begin{equation}
\Delta \wh{m^2}_{ee}=
\begin{cases}
\wh{m^2}_3-\wh{m^2}_1,\quad&a/\Delta m^2_{21}\ll-1\\
\wh{m^2}_3-\wh{m^2}_2,& a/\Delta m^2_{21}\gg1
\end{cases}\,,
\label{eq:ee0}
\end{equation}
where $a=2\sqrt2EG_FN_e$ is the matter potential, $G_F$ is Fermi's constant, $N_e$ is the electron density, and the $\wh{m^2}_i/2E$ are the exact eigenvalues which are calculated in \cite{Zaglauer:1988gz}, see also appendix \ref{sec:exact eigenvalues}.
This is independent of mass ordering.

We note that $\wh{m^2}_2$ and $\wh{m^2}_1$ are approximately constant for $a/\Delta m^2_{21} \ll -1$ and $a/\Delta m^2_{21} \gg 1$, respectively.
This suggests defining $\Delta \wh{m^2}_{ee} $ as follows\footnote{Note that $m^2_0$ is identical to $\lambda_b=\lambda_0$ from \cite{Denton:2016wmg}.}:
\begin{align}
\Delta \wh{m^2}_{ee}  &\equiv  \wh{m^2}_3-(\wh{m^2}_1 +\wh{m^2}_2 -m^2_0)\,, \label{eq:ee1}  \\[3mm]
{\rm where} \quad m^2_0 & \equiv \wh{m^2}_2(a=-\infty)=\wh{m^2}_1(a=+\infty) \nonumber \\
& =\Delta m^2_{21} c^2_{12}\,   \label{eq:msq0}
\end{align}
using the (convention dependent) asymptotic values for the eigenvalues shown in Table \ref{tab:1}. 
By construction, this reproduces eq.~\ref{eq:ee0}
for $|a/\Delta m^2_{21}|\gg1$ and is applicable for both mass orderings. The sign of $\Delta \wh{m^2}_{ee}$ determines the mass ordering.

It is also useful to note that $m^2_0$ can be written as 
\begin{equation}
 m^2_0  =  \Delta m^2_{ee}  -[m^2_3-(m^2_1 +m^2_2)]  \,.
 \label{eq:msq0defn}
\end{equation}
Then, as suggested by eq.~\ref{eq:ee-intro}, $\Delta\wh{m^2}_{ee}$ can also be written in the following simple and easy to remember form,
\begin{equation}
\Delta\wh{m^2}_{ee}-\dmsqee=(\wh{m^2}_3-m^2_3)-(\wh{m^2}_1-m^2_1)-(\wh{m^2}_2-m^2_2)\,,
\label{eq:differences}
\end{equation}
where recovery of the vacuum limit is manifest.
In the following sections we will address in more detail why the definition  of eq.\,\ref{eq:ee-intro} works for all matter potentials including $|a/\Delta m^2_{21}| \ll1$.

Here we will use eq.~\ref{eq:ee-intro} to re-write the $\wh{m^2}_i$'s in matter as a function of the two relevant $\Delta\wh{m^2}$'s: $\dmsqeea$ and $\Delta\wh{m^2}_{21}$.
By properties of the trace of the Hamiltonian\footnote{Explicitly, in the flavor basis we have that $2E\tr(H)=\tr(UMU^\dagger+A)=\tr(UU^\dagger M)+\tr(A)=\Delta m^2_{31}+\Delta m^2_{21}+a$. In the matter basis the trace of the Hamiltonian is $2E\tr(H)=\tr(\wh U\wh M\wh U^\dagger)=\tr(\wh U\wh U^\dagger\wh M)=\sum_i\wh{m^2}_i$.}, we have
\begin{equation}
\wh{m^2}_3+\wh{m^2}_2+\wh{m^2}_1 =\Delta m^2_{31}+\Delta m^2_{21} +a\,.
\end{equation}
Then together with eq.~\ref{eq:ee1} above
\begin{align}
\wh{m^2}_3 & = \Delta m^2_{31} +  \frac{1}{2}a +  \frac{1}{2}(\Delta \wh{m^2}_{ee}-  \Delta m^2_{ee} )\,, \nonumber\\
\wh{m^2}_2 + \wh{m^2}_1 & = \Delta m^2_{21} +  \frac{1}{2}a -  \frac{1}{2}(\Delta \wh{m^2}_{ee}-  \Delta m^2_{ee} )\,.\label{eq:msq321}
\end{align}
We make the typical definition $\Delta\wh{m^2}_{21}\equiv\wh{m^2}_2-\wh{m^2}_1$, then
\begin{align}
\wh{m^2}_1 ={}&\frac{1}{4}a - \frac{1}{4}(\Delta \wh{m^2}_{ee} - \Delta m^2_{ee})\nonumber\\
&-\frac{1}{2}(\Delta \wh{m^2}_{21}- \Delta m^2_{21} ) \nonumber \\
\wh{m^2}_2 ={}&\Delta m^2_{21} + \frac{1}{4}a - \frac{1}{4}(\Delta \wh{m^2}_{ee} - \Delta m^2_{ee})\nonumber\\
&+\frac{1}{2}(\Delta \wh{m^2}_{21}- \Delta m^2_{21} ) \nonumber \\
\wh{m^2}_3 ={}&\Delta m^2_{31} + \frac{1}{2}a + \frac{1}{2}(\Delta \wh{m^2}_{ee} - \Delta m^2_{ee})\,,\label{eq:msqi}
\end{align}
which implies
\begin{align}
\Delta \wh{m^2}_{31} ={}& \Delta m^2_{31} +  \frac{1}{4}a +  \frac{3}{4}(\Delta \wh{m^2}_{ee}  -  \Delta m^2_{ee}) \nonumber \\
&+\frac{1}{2}(\Delta \wh{m^2}_{21}- \Delta m^2_{21} )\label{eq:Dmsqi} \\
\Delta \wh{m^2}_{32} ={}& \Delta \wh{m^2}_{31} -  \Delta \wh{m^2}_{21}\,.\nonumber
\end{align}

\begin{table}
\renewcommand{\arraystretch}{1.2}
\centering
\hspace*{-3mm}
\begin{tabular}{|c|c|c|c|}
\hline
 & $a \rightarrow -\infty$ & $a=0$ &   $a \rightarrow +\infty$  \\\hline
 $  \wh{m^2}_3$ & $\Delta m^2_{ee} c^2_{13} + \Delta m^2_{21} s^2_{12}$ & $\Delta m^2_{31}$  & $ \,a +\Delta m^2_{ee} s^2_{13} + \Delta m^2_{21} s^2_{12}\,$ \\
$  \wh{m^2}_2$ & $\Delta m^2_{21} c^2_{12}$ &  $\Delta m^2_{21} $ & $\Delta m^2_{ee} c^2_{13} + \Delta m^2_{21} s^2_{12}$ \\
$  \wh{m^2}_1$ & $\,a +\Delta m^2_{ee} s^2_{13} + \Delta m^2_{21} s^2_{12}\,$ &  0 & $ \Delta m^2_{21} c^2_{12}$ \\\hline
\end{tabular}
\caption{The mass squareds in matter for various limits of $a$ in the NO.
See eqs.~5.3-5.4 of \cite{Minakata:2015gra} or Table 4 of \cite{Denton:2016wmg}. Adding the same constant to all entries in this table, does not effect oscillation physics. Our convention is that in vacuum $m^2_1= 0$.}
\label{tab:1}
\end{table}

We can also use $\Delta \wh{m^2}_{ee}$ to estimate  $ \Delta \wh{m^2}_{21}$ except near $a \approx 0$.  For $|a/\Delta m^2_{21}| \gg1$,
either $\wh{m^2}_2=m^2_0$  or  $\wh{m^2}_1=m^2_0$. 
Then, 
\begin{align}
\Delta \wh{m^2}_{21}  \approx{}&  | \wh{m^2}_2 + \wh{m^2}_1- 2m^2_0 | \nonumber \\
\approx{}&\Delta m^2_{21} \left|  \, a_{12}/ \Delta m^2_{21}- \cos 2 \theta_{12} \, \right|
\nonumber \\ &
+ {\cal O}(\Delta m^2_{21} )\,,
\label{eq:dmsq21_asym}
\end{align}
where we have made the natural definition,
\begin{equation}
a_{12} \equiv \frac{1}{2}(a+ \Delta m^2_{ee}- \Delta \wh{m^2}_{ee})
\end{equation}
as the effective matter potential for the 12 sector as was used in \cite{Denton:2018hal}.
For this derivation eq.~\ref{eq:msq321} is needed.

The asymptotic eigenvalues in Table \ref{tab:1}, can also be used to obtain a simple approximate expression for $ \Delta \wh{m^2}_{ee} $, when $|a| \gg \Delta m^2_{ee}$:
\begin{equation}
\Delta \wh{m^2}_{ee} \approx   \Delta m^2_{ee} \left| a/\Delta m^2_{ee}- \cos 2 \theta_{13}\right|.
\label{eq:dmsqee_asym}
\end{equation}

These two asymptotic expressions for  $\Delta \wh{m^2}_{ee}$ and $ \Delta \wh{m^2}_{21}$, eqs.~\ref{eq:dmsqee_asym} and \ref{eq:dmsq21_asym} respectively, which were obtained with only general information of the neutrino mass squareds in matter here, will be compared to the expressions obtained using the  approximations of \cite{Minakata:2015gra} \& \cite{Denton:2016wmg}  in  section \ref{sec:relation to DMP}.

\section{Three-flavor to Two-Flavor}
\label{sec:three to two}
Instead of studying the asymptotic behavior of $\Delta\wh{m^2}_{ee}$, we instead focus on explicitly connecting the three-flavor expression with the two-flavor expression.
The exact three-flavor $\nu_e$ disappearance probability in matter $P_a(\nu_e\to\nu_e)$ is given by
\begin{align}
1-P_a={}&4|\wh{U}_{e3}|^2\left[|\wh{U}_{e1}|^2 \sin^2 \wh{\Delta}_{31} + |\wh{U}_{e2}|^2 \sin^2 \wh{\Delta}_{32}\right]\nonumber\\
& + 4|\wh{U}_{e1}|^2|\wh{U}_{e2}|^2 \sin^2 \wh{\Delta}_{21} \nonumber\\
={}&\sin^2 2 \wh\theta_{13}\left[c^2_{\wh{12}} \sin^2 \wh{\Delta}_{31} + s^2_{\wh{12}} \sin^2 \wh{\Delta}_{32}\right]\nonumber\\
& + c^4_{\wh{13}} \sin^2 2 \wh{\theta}_{12} \sin^2 \wh{\Delta}_{21}\,, \label{eq:Pa3}
\end{align}
where we have used $s_{ij}=\sin\theta_{ij}$ and $c_{ij}=\cos\theta_{ij}$.
As was shown in \cite{Minakata:2007tn}, eq.~\ref{eq:Pa3} can be rewritten without approximation, as
\begin{multline}
1-P_a(\nu_e\to\nu_e) =c^4_{\wh{13}} \sin^2 2 \wh{\theta}_{12} \sin^2 \wh{\Delta}_{21}\\
+\frac12 \sin^2 2 \wh{\theta}_{13}  \left[1-\sqrt{1- \sin^2 2 \wh{\theta}_{12} \sin^2 \wh{\Delta}_{21}}\cos(2 \wh{\Delta}_{EE}+ \wh{\Omega} )\right]\,, 
\end{multline}
where $\wh{\Omega} = \arctan(\cos 2  \wh{\theta}_{12} \tan \wh{\Delta}_{21} ) - \wh{ \Delta}_{21} \cos 2  \wh{\theta}_{12}$ and
$\Delta\wh{m^2}_{EE}$ is a new frequency defined by
\begin{equation}
\Delta \wh{m^2}_{EE}\equiv\cos^2 \wh{\theta}_{12} \Delta \wh{m^2}_{31} +  \sin^2 \wh{\theta}_{12}\Delta \wh{m^2}_{32}\,.
\label{eq:EE}
\end{equation}

For $|E|$ greater than a few GeV, $\Delta \wh{m^2}_{21}\gg\Delta m^2_{21}$ (see fig.~\ref{fig:msqeds}) and therefore $ \wh{\theta}_{12} \approx 0$ or $\pi/2$, which makes  $ \sqrt{1-  \sin^2 2  \wh{\theta}_{12}   \sin^2 \wh{\Delta}_{21}} \approx 1$ and $\wh{\Omega}\approx 0$. Hence, 
\begin{equation*}
1-P_a(\nu_e\to\nu_e) \approx \sin^2 2 \wh{\theta}_{13} \sin^2 \wh{\Delta}_{EE}\,,
\end{equation*}
in agreement with eq.~\ref{eq:Pa2} in this energy range\footnote{Note $\sin^2 2 \wh{\theta}_{13} > \wh{c}^{\,4}_{13} \sin^2 2 \wh{\theta}_{12}$ except when $|E| < 1.1$ GeV, see fig.~\ref{fig:R2}. We take $\rho=3$ g/cc throughout the article.}. 
Also in this energy region, it is clear that\footnote{This statement is made under the assumption that $\wh\theta_{12}\to\pi/2$ $(0)$ as $a\to\infty$ $(-\infty)$.  In fact, there is a small correction to this assumption. In this limit, $\sin^2\wh\theta_{12}=1-{\cal O}(\eps^{\prime2})$ where $\eps^{\prime2}<3\e{-4}$, \cite{Denton:2018fex}.}
\begin{equation}
\Delta \wh{m^2}_{EE} \approx
\begin{cases}
\Delta \wh{m^2}_{31},  \quad & a \ll \Delta m^2_{21} \\
\Delta \wh{m^2}_{32},  \quad & a \gg \Delta m^2_{21} \,.
\end{cases}
\end{equation}

Using the explicit results from \cite{Zaglauer:1988gz}, it is simple to show, without approximation, that
\begin{equation}
\Delta \wh{m^2}_{EE} = \frac{(\wh{m^2}_3-\wh{m^2}_a)(\wh{m^2}_3-\wh{m^2}_1)(\wh{m^2}_3-\wh{m^2}_2)}{
(\wh{m^2}_3)^2- \wh{m^2}_3\wh{m^2}_a - \beta +\wh{m^2}_1\wh{m^2}_2}\,,
\end{equation}
where 
\begin{align*}
\beta  & \equiv \Delta m^2_{ee} c^2_{13} \Delta m^2_{21} c^2_{12}=   \wh{m^2}_1\wh{m^2}_2 \wh{m^2}_3 /a \\
\wh{m^2}_a & \equiv  a +\Delta m^2_{ee} s^2_{13} +\Delta m^2_{21} s^2_{12}\,. 
\end{align*}
Note\footnote{Also note that $\wh{m^2}_a$ is identical to $\lambda_a$ from \cite{Denton:2016wmg}.} that $\wh{m^2}_3(a \rightarrow \infty)  \rightarrow    \wh{m^2}_a$ and $\wh{m^2}_1(a \rightarrow - \infty)  \rightarrow   \wh{m^2}_a$.

In the low energy limit, when $|\wh{m^2}_3| \gg |\wh{m^2}_j|$ for $j=(1, 2, a)$, a first order perturbative expansion in $\wh{m^2}_j/\wh{m^2}_3$ gives
\begin{equation}
\Delta \wh{m^2}_{EE}\approx \wh{m^2}_3-( \wh{m^2}_1+ \wh{m^2}_2- m^2_0),
\end{equation}
consistent with our previous definition, eq.~\ref{eq:ee1}.  In fact, $\Delta \wh{m^2}_{ee}$ and $\Delta \wh{m^2}_{EE}$ differ by less than  $<$ 0.3\% for all values of matter potential. 

In vacuum ($E=0$), it is known that eq.~\ref{eq:P0} is an excellent approximation over the first couple of oscillations see e.g.~\cite{Parke:2016joa}, further verifying the use of this two-flavor approximation. The analysis  of this paper can be trivially extend away from vacuum region using the matter oscillation parameters.

\section{Relation to DMP approximation}
\label{sec:relation to DMP}
While eq.~\ref{eq:ee1} is a compact expression that behaves as we expect $\dmsqeea$ ought to, it is not simple due to the complicated expressions for the eigenvalues, in particular the $\cos(\frac13\cos^{-1}\dots)$ part of each eigenvalue, see appendix \ref{sec:exact eigenvalues}.
In order to both verify the behavior of $\dmsqeea$ for $|a/\Delta m^2_{ee}|\ll1$ and provide an expression that is simple we look to approximate expressions of the eigenvalues.

In refs.~\cite{Minakata:2015gra}, \cite{Denton:2016wmg} \& \cite{Denton:2018hal} (DMP) simple, approximate, and precise analytic expressions were given for neutrino oscillations in matter.
In the DMP approximation\footnote{In the notation of DMP, $\Delta \wt{m^2}_{ee} \equiv \Delta \lambda_{+-} = \cos^2 \psi \, \Delta \lambda_{31} + \sin^2 \psi  \, \Delta \lambda_{32}$, see eq.~A.1.7 of \cite{Denton:2016wmg}.
Also, $\wt\theta_{12}=\psi$ and $\wt{m^2}_i=\lambda_i$ in DMP; see \cite{Denton:2018hal}.} through zeroth order, the definition of $\Delta \wh{m^2}_{ee} $ given in eq.~\ref{eq:ee1} can be shown to be
\begin{align}
\Delta \wh{m^2}_{ee} & \approx \wt{m^2}_3-(\wt{m^2}_1 +\wt{m^2}_2 -m^2_0)\equiv \Delta \wt{m^2}_{ee} \,,  \nonumber \\
& =\cos^2 \wt\theta_{12} \Delta \wt{m^2}_{31} + \sin^2 \wt\theta_{12} \Delta \wt{m^2}_{32}\,,\label{eq:dmp}\\
& = \Delta m^2_{ee} \sqrt{(\cos 2\theta_{13} -a/\Delta m^2_{ee})^2+ \sin^2 2\theta_{13}}\,,\nonumber 
\end{align}
where $\wt\theta_{12}$ and $\wt\theta_{13}$ are excellent approximations for the matter mixing angles $\wh\theta_{12}$ and $\wh\theta_{13}$ and $\Delta\wt{m^2}_{31}$ and $\Delta\wt{m^2}_{32}$ are the corresponding approximate expressions for $\Delta \wh{m^2}_{31}$ and $\Delta\wh{m^2}_{31}$ from \cite{Denton:2016wmg} and reproduced in appendix \ref{sec:DMP} below\footnote{The notation is such that while both $\wh x$ and $\wt x$ are quantities in matter, $\wh x$ denotes the exact quantity and $\wt x$ denotes the zeroth order approximation from DMP, and $\wt x$ is an excellent approximation for $\wh x$.}.
The approximation has corrections to the eigenvalues of $\mathcal O(\eps^{\prime2})$ where $\eps'=\sin(\wt\theta_{13}-\theta_{13})s_{12}c_{12}\Delta m^2_{21}/\Delta m^2_{ee}$.
$|\eps'|<0.015$ and is equal to zero in vacuum. Equation \ref{eq:dmp} provides a very simple means to modify the vacuum $\Delta m^2_{ee}$ to get the corresponding expression in matter.

In the DMP approximation, all three expressions, eq.~\ref{eq:dmp}, for  $\Delta \wt{m^2}_{ee}$ can be shown to be analytically identical. This is however not true for the exact eigenvalues and mixing angles in matter, there are small differences between these expressions (quote fractional differences.). We use the first line of eq.~\ref{eq:dmp} for our definition $\Delta {m^2}_{ee}$ in matter, because this definition allows us a general understanding of the three neutrino eigenvalues in matter (see eqs.~\ref{eq:msqi} and \ref{eq:Dmsqi}).
We now verify that this definition of $\Delta {m^2}_{ee}$ in matter meets all the other criteria we need it to.

First we see that by using the DMP zeroth order approximation, $\Delta \wt{m^2}_{ee}$ is just the matter generalization of the vacuum expression,  $\Delta  m^2_{ee}= \cos^2 \theta_{12} \Delta m^2_{ee} +\sin^2 \Delta m^2_{32}$ and provides a connection to why the definition of eq.~\ref{eq:ee1} works for $|a/\Delta m^2_{21}| < 1$ also.
 
Asymptotically, as $|a/\Delta m^2_{ee}| \gg 1$, in this approximation scheme
\begin{equation}
\Delta \wt{m^2}_{ee} \rightarrow   \Delta m^2_{ee} \left|a/ \Delta m^2_{ee} -\cos 2\theta_{13}\right|\,,
\label{eq:dmpinfty}
\end{equation}
in agreement with eq.~\ref{eq:dmsqee_asym}.

Similarly for $\Delta \wt{m^2}_{21}$, from DMP
\begin{multline}
\Delta\wt{m^2}_{21} = \Delta m^2_{21}\left[(\cos 2\theta_{12} - \wt{a}_{12} /\Delta m^2_{21})^2 \vphantom{\wt{\theta}_{13}}\right.\\
\left.+\sin^2 2 \theta_{12} \cos^2(\wt{\theta}_{13}-\theta_{13})\right]^{1/2}\,,
\label{eq:dmsq21}
\end{multline}
where $\wt{a}_{12} \equiv (a+\Delta m^2_{ee}-\Delta \wt{m^2}_{ee})/2$ and
\begin{equation}
\cos^2(\wt{\theta}_{13}-\theta_{13}) = \frac{\Delta \wt{m^2}_{ee}+\Delta m^2_{ee}-a \cos 2\theta_{13}}{2\Delta \wt{m^2}_{ee}}\,.
\end{equation}
Asymptotically,  $|a/\Delta m^2_{21}| \gg 1$, we have
\begin{equation}
\Delta \wt{m^2}_{21} \rightarrow \left|\Delta m^2_{21} \cos 2 \theta_{12} 
   -  \frac{1}{2}\left(a+\Delta m^2_{ee}  -\Delta \wt{m^2}_{ee}\right)\right|\,,
\end{equation}
again in agreement with eq.~\ref{eq:dmsq21_asym}.  So everything discussed in section \ref{sec:defn} is consistent with the simple and compact DMP approximation.  

In the next section we will analytically and then numerically show that the fractional difference between the two expressions, $\Delta \wt{m^2}_{ee}$ and $\Delta \wt{m^2}_{EE}$, are small.

\section{Comparison of the Two Expressions}
\label{sec:two expressions}
As previously shown the vacuum $\dmsqee$ can be written in two equivalent ways,
\begin{align*}
\dmsqee&=c_{12}^2\Delta m^2_{31}+s_{12}^2\Delta m^2_{32}\,,\\
&=m_3^2-m_1^2-m_2^2+m_0^2\,.
\end{align*}
The two expressions can be seen as two choices for the how to relate these to the matter version: one is to elevate each eigenvalue to its matter equivalent (everything except $m_0^2$) and the other is to elevate each term including the mixing angles.
We refer to the former as $\Delta\wh{m^2}_{ee}$ and the latter as $\Delta\wh{m^2}_{EE}$.

To understand how these expressions differ, we carefully examine their difference,
\begin{equation}
\Delta_{Ee}\equiv\Delta\wh{m^2}_{EE}-\Delta\wh{m^2}_{ee}=\wh{m_1^2}+c_{\wh{12}}^2\Delta\wh{m^2}_{21}-c_{12}^2\Delta m^2_{21}\,.
\end{equation}
We now quantify the difference between these expressions using DMP.
If both expressions provide good approximations for the two flavor frequency in matter then the difference between them should be small.
At zeroth order the difference is
\begin{equation}
\Delta_{Ee}^{(0)}=\wt{m^2}_1+c_{\wt{12}}^2\Delta\wt{m^2}_{21}-c_{12}^2\Delta m^2_{21}=0\,,
\end{equation}
so these expressions are exactly equivalent at zeroth order.

At first order the eigenvalues receive no correction, but $\wt\theta_{12}$ does.
From \cite{Denton:2018fex} we have that the first order correction is
\begin{equation}
\wt\theta_{12}^{(1)}=-\eps'\dmsqee t_{\wt{13}}\left(\frac{s_{\wt{12}}^2}{\Delta\wt{m^2}_{31}}+\frac{c_{\wt{12}}^2}{\Delta\wt{m^2}_{32}}\right)\,,
\end{equation}
where $t_{ij}=\tan\theta_{ij}$.
This leads to a correction of,
\begin{equation}
\Delta_{Ee}^{(1)}=t_{\wt{13}}s_{12}^2c_{12}^2\sin2\theta_{13}a\frac{(\Delta m^2_{21})^2}{\Delta\wt{m^2}_{32}\Delta\wt{m^2}_{31}}\,.
\end{equation}
As expected $\Delta_{Ee}\propto a$ for small $a$.
Also, we can verify that $\Delta_{Ee}/\Delta\wh{m^2}_{ee}$ is always small by seeing that $a/\Delta\wh{m^2}_{ee}$ remains finite and the only case where $t_{\wt{13}}\propto a$ for $a\to\infty$, but $\Delta\wt{m^2}_{32}\Delta\wt{m^2}_{31}\propto a^2$, thus the difference between the two expressions is always small.

\begin{figure}
\centering
\includegraphics[width=\columnwidth]{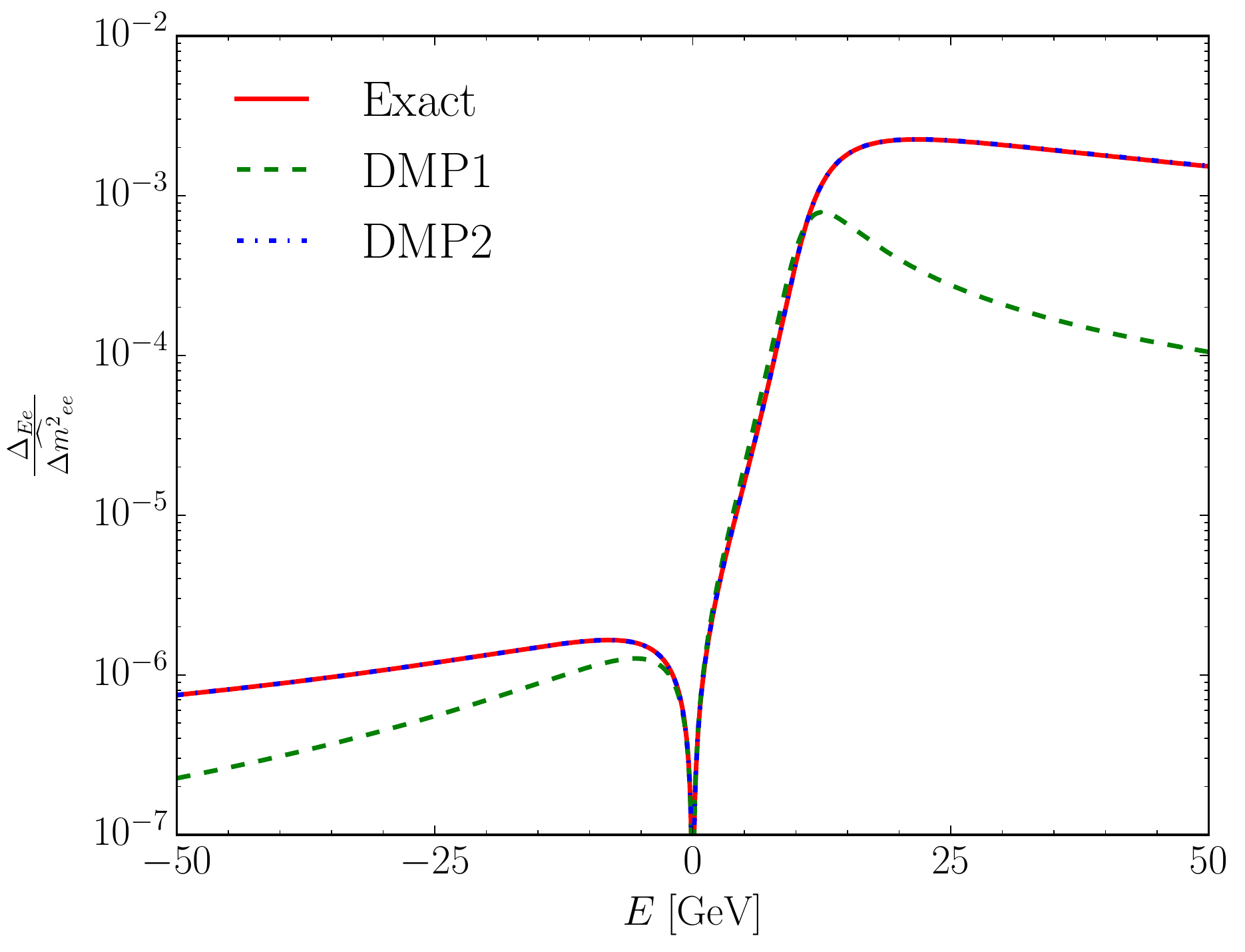}
\caption{The fractional difference between the two expressions is shown in the red solid curve.
The green dashed curve shows the difference through first order, and the blue dash-dotted curve shows the difference through second order.
Note that at zeroth order in DMP the difference is exactly zero. DMP2 is hard to see at it is on top of exact.}
\label{fig:AvB}
\end{figure}

$\Delta_{Ee}^{(1)}$ provides an adequate approximation of the difference between $\Delta\wh{m^2}_{ee}$ and $\Delta\wh{m^2}_{EE}$ as shown in fig.~\ref{fig:AvB}.
A precise estimate of the difference requires the second order correction to $\wt{\theta}_{12}$ given explicitly in \cite{Denton:2018fex} along with the second order corrections to the eigenvalues from DMP.
This is because this difference $\Delta_{Ee}$ depends strongly on the asymptotic behavior of $\wt\theta_{12}$ which only becomes precise beyond the atmospheric resonance at second order.
The result of this is also shown in fig.~\ref{fig:AvB} which shows that first order is not sufficient to accurately describe the difference, but second order is.
We see that for neutrinos the expressions agree to $\lesssim0.3\%$, and the agreement is $\sim3$ orders of magnitude better for anti-neutrinos.

In the next section we will investigate how well the two-flavor approximation, eq.~\ref{eq:Pa2}, works numerically for both the depth and position over the first oscillation minimum for $\nu_e$ disappearance for all values of the neutrino energy.

\section{Precision Verification}
\label{sec:precision}
The goal of $\dmsqeea$ is to provide the correct frequency such that the two-flavor disappearance expression, eq.~\ref{eq:Pa2}, is an excellent approximation for $\nu_e$ disappearance over the first oscillation in matter.
In particular, we want this expression to reproduce the position and depth of the first oscillation minimum at high $E$ (small $L$) correctly compared to the complete three-flavor picture.

\subsection{Numerical Comparison}
\begin{figure}
\centering
\includegraphics[width=\columnwidth]{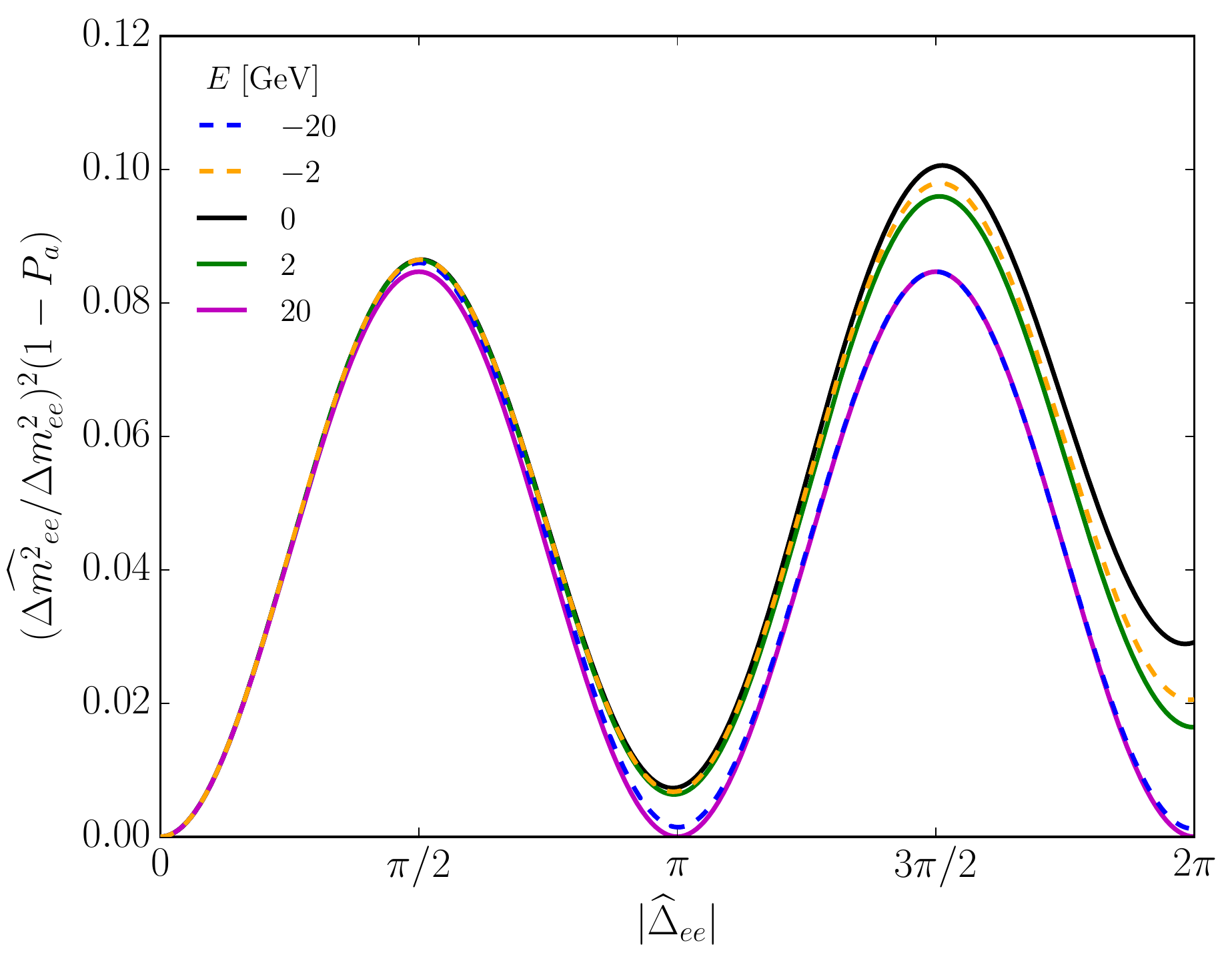}
\caption{Here we demonstrate the validity of the two-flavor approximation by plotting eq.~\ref{eq:fig2} showing the expected sinusoidal dependence. Here $P_a$ is the exact three flavor $\nu_e$ disappearance probability. Note the small deviations due to the 21 term that grow as the phase $|\widehat\Delta_{ee}|$ increases for small energies.}
\label{fig:Einvar}
\end{figure}

Using the definition of $\dmsqeea$ given in eq.~\ref{eq:ee1}, we plot in fig.~\ref{fig:Einvar}
\begin{equation}
\left(\frac{\Delta \wh{m^2}_{ee}}{\Delta m^2_{ee} } \right)^2  (1- P_a(\nu_e\to\nu_e)) 
\quad   {\rm verses} \quad  \wh{\Delta}_{ee}\,,
\label{eq:fig2}
\end{equation}
for various values of the neutrino energy.
Here $P_a(\nu_e\to\nu_e)$ is evaluated using the exact oscillation probability given in \cite{Zaglauer:1988gz}.
We see that this behaves like $\sin^2\wh\Delta_{ee}$ as expected, with increasing precision for increasing energy.
Note the approximate neutrino energy independence of this figure, demonstrating the universal nature of the approximation given in eq.~\ref{eq:Pa2} using our definition of $\dmsqeea$.

Next, we want to check that this two-flavor expression reproduces the first oscillation minimum at high $E$ (small $L$) correctly compared to the complete three-flavor picture.
The minimum occurs when the derivate of $P$ is zero.
We now have a choice: we can define the minimum when $dP_a/dL=0$ or $dP_a/dE=0$.
Since both $\wh\theta$ and $\dmsqeea$ are nontrivial functions of $E$, the correct option is to use $dP_a/dL=0$.

In order to numerically test the various expressions, we find the location $L$ of the first minimum by solving $dP_a/dL=0$ for a given $E$ using the full three-flavor expressions.
We then convert the $(L,E)$ pair at the first minimum into the corresponding $\dmsqeea$ using
\begin{equation}
\frac{\dmsqeea L}{4E}=\frac\pi2\,.
\label{eq:dmsqee LE}
\end{equation}
Next, we compare the difference between this numeric solution and the expressions presented in this paper, eqs.~\ref{eq:ee-intro}, \ref{eq:EE}, and \ref{eq:dmp}.
We also compare to the approximate analytic solution from \cite{Minakata:2017ahk} (HM), see appendix \ref{sec:HM}.
This comparison is shown in fig.~\ref{fig:Error_no21}.

\begin{figure}
\centering
\includegraphics[width=\columnwidth]{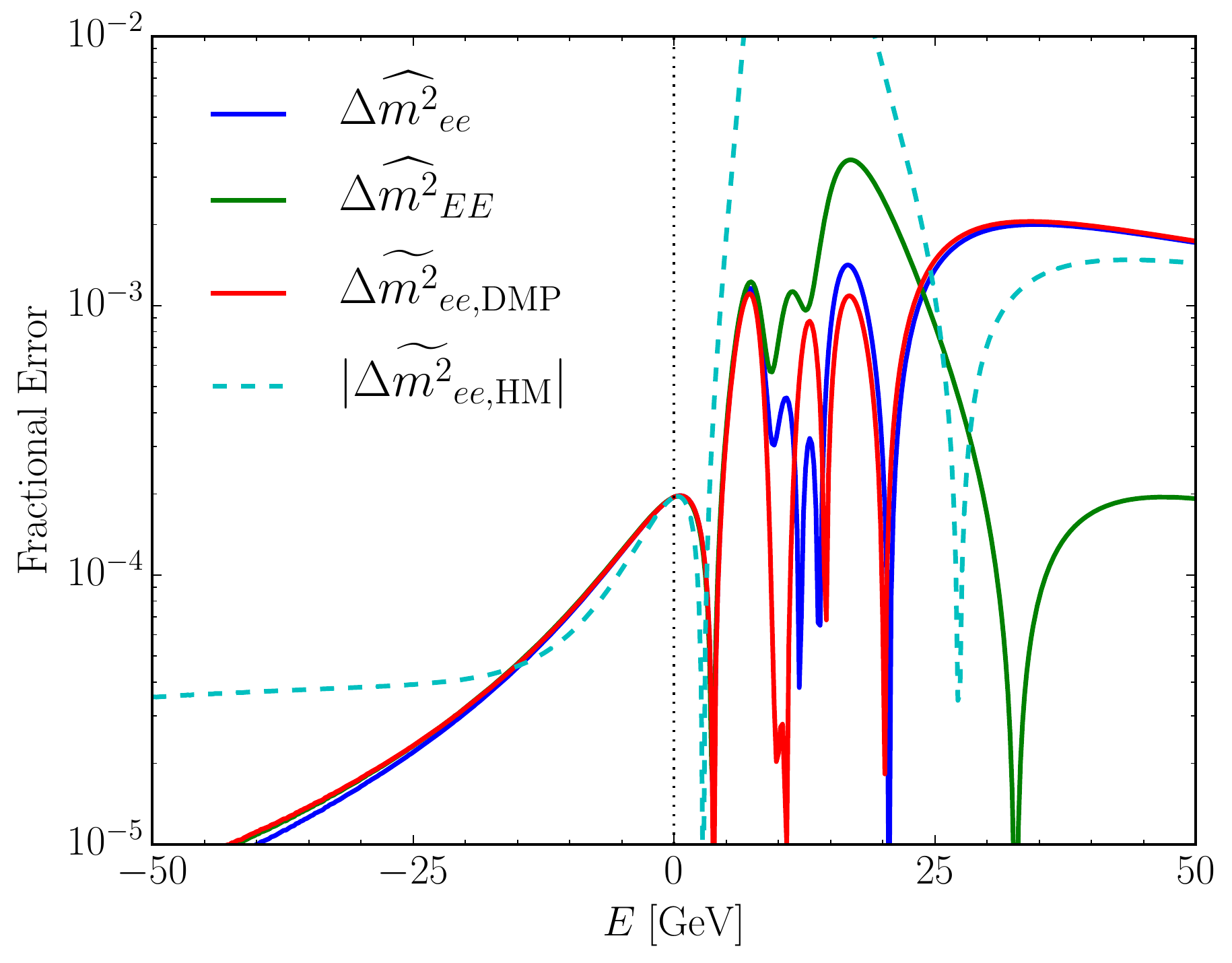}
\caption{We show the fractional error ($\delta x/x$) of various different $\Delta\widehat{m^2}_{ee}$ expressions with the precise numerical one determined at the point where $dP_a/dL=0$, see eq.~\ref{eq:dmsqee LE}.
For the exact numerical expression we ignore the $\Delta\widehat{m^2}_{21}$ term as no definition will get it correct.
The $ee$ curve uses the formula from eq.~\ref{eq:ee-intro} and the $EE$ curve uses the formula from eq.~\ref{eq:EE}.
The DMP curve uses the zeroth order expressions \cite{Denton:2016wmg} in the same formula which leads to the simple expression shown in eq.~\ref{eq:dmp}.
The HM curve uses the expression from \cite{Minakata:2017ahk} and takes the absolute value to get the sign correct for large $E$, see appendix \ref{sec:HM}.
We have fixed $\rho=3$ g/cc and assumed the NO.
$E>0$ corresponds to neutrinos, $E<0$ corresponds to anti-neutrinos, and $E=0$ corresponds to the vacuum.}
\label{fig:Error_no21}
\end{figure}

When determining the minimum from the exact expression, a two-flavor expression using only $\dmsqeea$ will get the $\Delta m^2_{31}$ and $\Delta m^2_{32}$ terms correct including matter effect, but will always be off by $\Delta m^2_{21}$ terms.
Thus in fig.~\ref{fig:Error_no21} we don't include the effect of the 21 term which will affect any two-flavor approximation comparably.

We see that for either eq.~\ref{eq:ee1} or eq.~\ref{eq:dmp} the agreement is excellent with relative error $<0.2\%$.
In addition, the two expressions clearly agree with each other to a higher level of precision than is necessary.
For the HM expression the agreement is good for anti-neutrinos and in the high energy limit, but is poor in a broad range near the atmospheric resonance for neutrinos.
In addition, we have modified the HM expression by taking the absolute value so that the HM expression asymptotically returns to the correct expression past the atmospheric resonance for neutrinos.

We have also compared $\dmsqeea$ with the exact solution including the $\Delta m^2_{21}$ term and found agreement to better than 1\%.

\subsection{Analytic Comparison}
\label{ssec:analytic comparison}
We now analytically estimate the precision of the two-flavor expression, for both the small $E$ (large $L$) limit and the large $E$ (small $L$) limit.

First, if $\Delta\wh{m^2}_{21}\ll|\Delta\wh{m^2}_{ee}|$ then at the $n^{\text{th}}$ oscillation minimum the ratio of the 21 term to the $ee$ term is well approximated by
\begin{equation}
\frac{\Delta m^2_{21}}{\dmsqee}[(2n-1)\pi/4]^2\label{eq:R}\,,
\end{equation}
as derived in appendix \ref{sec:ranges}.
For the first (second) oscillation peak this yields an error estimate of $<2\%$ (16\%); this two-flavor approach breaks down for $n>5$ when the ratio is $>1$.

The second case is when $\Delta\wh{m^2}_{21}\simeq|\Delta\wh{m^2}_{ee}|$, which occurs away from vacuum (high $E$, low $L$), and the ratio of the 21 coefficient to the $ee$ coefficient is
\begin{equation}
\frac{c_{\wh{13}}^4\sin^22\wh\theta_{12}}{\sin^22\wh\theta_{13}}=\frac{|\wh{U}_{e1}|^2|\wh{U}_{e2}|^2}{|\wh{U}_{e3}|^2(1-|\wh{U}_{e3}|^2)}\,,
\label{eq:R2}
\end{equation}
which is small away from vacuum as desired.
In particular, it is $<1$ for $|E|>1$ GeV.
See appendix \ref{sec:ranges} for details and numerical confirmation of each region.

\section{Conclusions}
\label{sec:conclusions}
In this paper, we have demonstrated that
\begin{align}
\Delta \wh{m^2}_{ee} & \equiv \wh{m^2}_3-(\wh{m^2}_1 +\wh{m^2}_2) \nonumber \\
 &  -[m^2_3-(m^2_1 +m^2_2)]  + \Delta m^2_{ee}  \\
& \approx  \Delta m^2_{ee} \sqrt{(\cos 2\theta_{13} -a/\Delta m^2_{ee})^2+ \sin^2 2\theta_{13}}\,,\nonumber
\end{align}
is the matter generalization of vacuum $\Delta m^2_{ee}$ that has been widely used by the short baseline reactor experiments Daya Bay and RENO and will be precisely measured ($<$ 1\%) in the medium baseline JUNO experiment. 
The exact and approximate expressions in the above equation  differ by no more than 0.06\%.
Another natural choice called $\Delta\wh{m^2}_{EE}$ is numerically very close to $\Delta\wh{m^2}_{ee}$ but does not provide the ability to simply rewrite the eigenvalues as $\Delta\wh{m^2}_{ee}$ does.

For $\nu_e$ disappearance in matter the position of the first oscillation minimum, for fixed neutrino energy E, is given by
\begin{equation}
L = \frac{2 \pi E}{\Delta \wh{m^2}_{ee}}\,,
\end{equation}
and the depth of the minimum is controlled by
\begin{align}
\sin^2 2 \wh\theta_{13} & \approx  \sin^2 2 \theta_{13} \left( \frac{\Delta m^2_{ee}}{\Delta \wh{m^2}_{ee}} \right)^2\,, \label{eq:depth}  
\\[2mm]
&\approx\frac{\sin^22\theta_{13}}{(\cos^22\theta_{13}-a/\dmsqee)^2+\sin^22\theta_{13}}\,.\nonumber
\end{align}
This two-flavor approximate expression is not only simple and compact, but it is precise to within $<1\%$ precision at the first oscillation minimum\footnote{In eq.~\ref{eq:depth}, the exact and second approximation differ in value by no more than $4 \times 10^{-4}$ and the fractional difference is smaller than 0.1\% except for very large positive values of the energy where the fractional difference is however never larger than 1\%.}.

The combination of $ \Delta \wh{m^2}_{ee}$ and $ \Delta \wh{m^2}_{21}$ is very powerful for understanding the effects of matter on the eigenvalues and  the mixing angles of the neutrinos. 
In this article we have illuminated the exact nature of $\Delta \wh{m^2}_{ee}$ and $\Delta\wh{m^2}_{21}$ which were extensively used in DMP \cite{Denton:2016wmg,Denton:2018hal}.

\begin{acknowledgments}
We thank Hisakazu Minakata for comments on an earlier version of this manuscript.

This manuscript has been authored by Fermi Research Alliance, LLC under Contract No.~DE-AC02-07CH11359 with the U.S. Department of Energy, Office of Science, Office of High Energy Physics.

This project has received funding/support from the European Union's Horizon 2020 research and innovation programme under the Marie Sklodowska-Curie grant agreement No 690575 \& No 674896.

PBD acknowledges support from the Villum Foundation (Project No.~13164) and the Danish National Research Foundation (DNRF91 and Grant No.~1041811001).
\end{acknowledgments}

\appendix
\section{Exact Eigenvalues}
\label{sec:exact eigenvalues}
From \cite{Zaglauer:1988gz} the exact eigenvalues in matter are $\wh{m^2}_i/2E$ where the $\wh{m^2}_i$ are
\begin{align}
\wh{m^2}_1&=\frac w3-\frac13 z \sqrt{w^2-3x}-\frac1{\sqrt3} \sqrt{1-z^2} \sqrt{w^2-3x}\,,\nonumber\\
\wh{m^2}_2&=\frac w3-\frac13 z \sqrt{w^2-3x}+\frac1{\sqrt3}\sqrt{1-z^2} \sqrt{w^2-3x}\,,\nonumber\\
\wh{m^2}_3&=\frac w3+\frac23 z \sqrt{w^2-3x}\,,\label{eq:ZSmsqs}
\end{align}
where
\begin{align}
w&=\Delta m^2_{21}+\Delta m^2_{31}+a\,,\nonumber\\
x&=\Delta m^2_{31}\Delta m^2_{21}+a\left[\Delta m^2_{31}c_{13}^2+\Delta m^2_{21}(c_{13}^2c_{12}^2+s_{13}^2)\right]\,,\nonumber\\
y&=a\Delta m^2_{31}\Delta m^2_{21}c_{31}^2c_{12}^2\,, \nonumber \\
z&=\cos\left\{\frac{1}{3} \cos^{-1}\left[\frac{2w^3-9wx+27y}{2(w^2-3x)^{3/2}} \right]\right\}\,.
\end{align}

Therefore,
\begin{align}
\Delta \wh{m^2}_{ee} &=\frac43 z \sqrt{w^2-3x} - \frac w3 +\Delta m^2_{21} c^2_{12}\, ,\nonumber\\
\Delta \wh{m^2}_{21}&=\frac2{\sqrt3}\sqrt{1-z^2} \sqrt{w^2-3x}\,.\label{eq:ZSDmsqs}
\end{align}
Using eq.~\ref{eq:ZSDmsqs} in eq.~\ref{eq:ZSmsqs} reproduces eq.~\ref{eq:msqi}, as a cross check.

\section{DMP Approximate Expression}
\label{sec:DMP}
Here we review the approximate expressions for the mixing angles and eigenvalues derived in \cite{Denton:2016wmg}.
The result of the 13 rotation yields
\begin{equation}
\Delta\wt{m^2}_{ee}=\dmsqee\sqrt{(\cos2\theta_{13}-a/\dmsqee)^2+\sin^22\theta_{13}}\,,
\label{eq:DMP Dmsqeea}
\end{equation}
\begin{equation}
\cos2\wt\theta_{13}=\frac{\dmsqee\cos2\theta_{13}-a}{\Delta\wt{m^2}_{ee}}\,.
\end{equation}
The 21 rotation yields
\begin{multline}
\Delta\wt{m^2}_{21}=\Delta m^2_{21}\left[(\cos2\theta_{12}-a_{12}/\Delta m^2_{21})^2\vphantom{\wt\theta_{13}}\right.\\
\left.+\cos^2(\wt\theta_{13}-\theta_{13})\sin^22\theta_{12}\right]^{1/2}\,,
\label{eq:DMP Dmsq21a}
\end{multline}
\begin{equation}
\cos2\wt\theta_{12}=\frac{\Delta m^2_{21}\cos2\theta_{12}-\wt{a}_{12}}{\Delta\wt{m^2}_{21}}\,,
\end{equation}
where we similarly define $\wt{a}_{12}\equiv(a+\dmsqee-\Delta\wt{m^2}_{ee})/2$.
Finally, from eqs.~\ref{eq:DMP Dmsqeea} and \ref{eq:DMP Dmsq21a} it is straightforward to show that
\begin{multline}
\Delta\wt{m^2}_{31}=\Delta m^2_{31}+\frac14a+\frac12(\Delta\wt{m^2}_{21}-\Delta m^2_{21})\\+\frac34(\Delta\wt{m^2}_{ee}-\dmsqee)\,.
\end{multline}
The remaining two oscillation parameters, $\wt\theta_{23}=\theta_{23}$ and $\wt\delta=\delta$, remain unchanged in this approximation.
We note that for each parameter above $\wt{x}$ provides an excellent approximation for $\wh{x}$.

We also note two additional useful expressions,
\begin{align}
\sin2\wt\theta_{13}&=\sin2\theta_{13}\left(\frac{\dmsqee}{\Delta\wt{m^2}_{ee}}\right)\,,\label{eq:s213m}\\
\sin2\wt\theta_{12}&=\cos(\wt\theta_{12}-\theta_{12})\sin2\theta_{12}\left(\frac{\Delta m^2_{21}}{\Delta\wt{m^2}_{21}}\right)\,.\label{eq:s212m}
\end{align}

\section{Alternate Expression}
\label{sec:HM}
An alternate approximate expression was previously provided in \cite{Minakata:2017ahk}, the expression from that paper is
\begin{multline}
\Delta\wt{m^2}_{ee,{\rm HM}}=(1-r_A)\dmsqee\\+r_A\left(\frac{2s_{13}^2}{1-r_A}\Delta m^2_{31}-s_{12}^2\Delta m^2_{21}\right)\,,
\label{eq:HMDmsqeea}
\end{multline}
where $r_A\equiv a/\Delta m^2_{31}$.
This expression clearly has a pole at $a=\Delta m^2_{31}$ which is the atmospheric resonance for neutrinos.
In addition, past the resonance, for $a>\Delta m^2_{31}$, the sign is incorrect as $\Delta\wt{m^2}_{ee,{\rm HM}}<0$ for the NO.
Thus we take the absolute value in our numerical studies.

In fig.~2 of \cite{Minakata:2017ahk}, the author compared eq.~\ref{eq:HMDmsqeea} with the minimum  obtained via solving $dP_a/dE=0$ whereas we have argued in section \ref{sec:precision} that a better comparison is obtained by solving $dP_a/dL=0$ for fixed E.

\section{Precision in Different Ranges}
\label{sec:ranges}
In this appendix we further expand upon the discussion in subsection \ref{ssec:analytic comparison}.

The exact three-flavor expression in matter from eq.~\ref{eq:Pa3} can be written as,
\begin{align}
1-P_a={}&\sin^22\theta_{13}\left(\frac{\dmsqee}{\dmsqeea}\right)^2\sin^2\wh\Delta_{ee}\nonumber\\
&+C(E)c_{\wh{13}}^4\sin^22\wh\theta_{12}\sin^2\wh\Delta_{21}\,,\label{eq:PC}
\end{align}
where $C(E)\simeq1$ contains the correction between the first and second term.
For the two-flavor approximation to be valid, the 21 term, $C(E)c_{\wh{13}}^4\sin^22\wh\theta_{12}\sin^2\wh\Delta_{21}$ must be small compared to the two-flavor $ee$ term, $\sin^22\wh\theta_{13}\sin^2\wh\Delta_{ee}$.
As in section \ref{ssec:analytic comparison}, we consider two cases.

\begin{figure}
\centering
\includegraphics[width=\columnwidth]{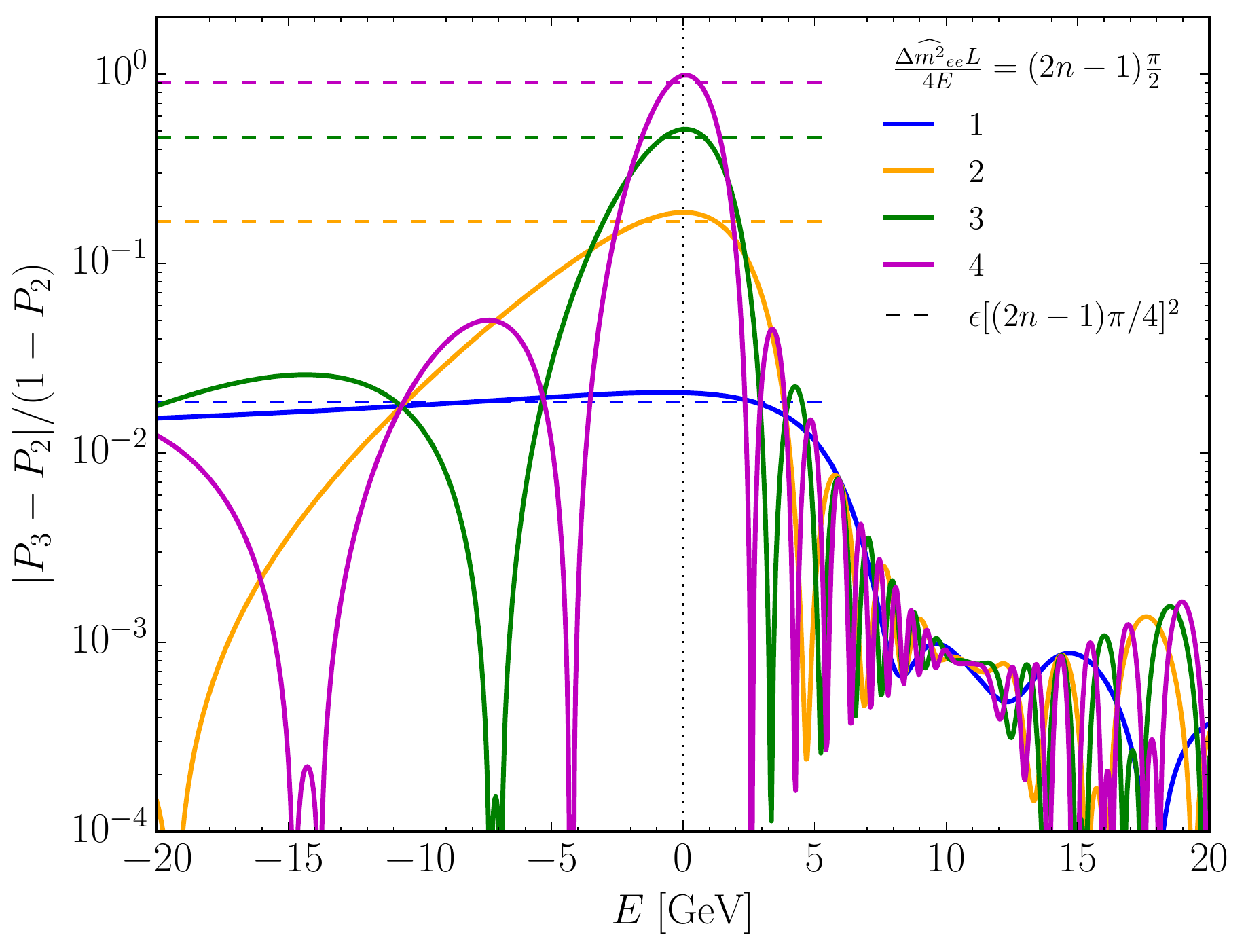}
\caption{The error of the two-flavor approximation ($P_2$ from eq.~\ref{eq:Pa2}) compared to the full three-flavor expression ($P_3$ from eq.~\ref{eq:Pa3}) in matter is shown in the solid curves for the first several oscillation minima.
The dashed lines are the simple approximation from eq.~\ref{eq:R}.
As expected eq.~\ref{eq:R} performs well near vacuum at $|E|\lesssim$ few GeV.}
\label{fig:Error_21}
\end{figure}

\begin{figure}
\centering
\includegraphics[width=\columnwidth]{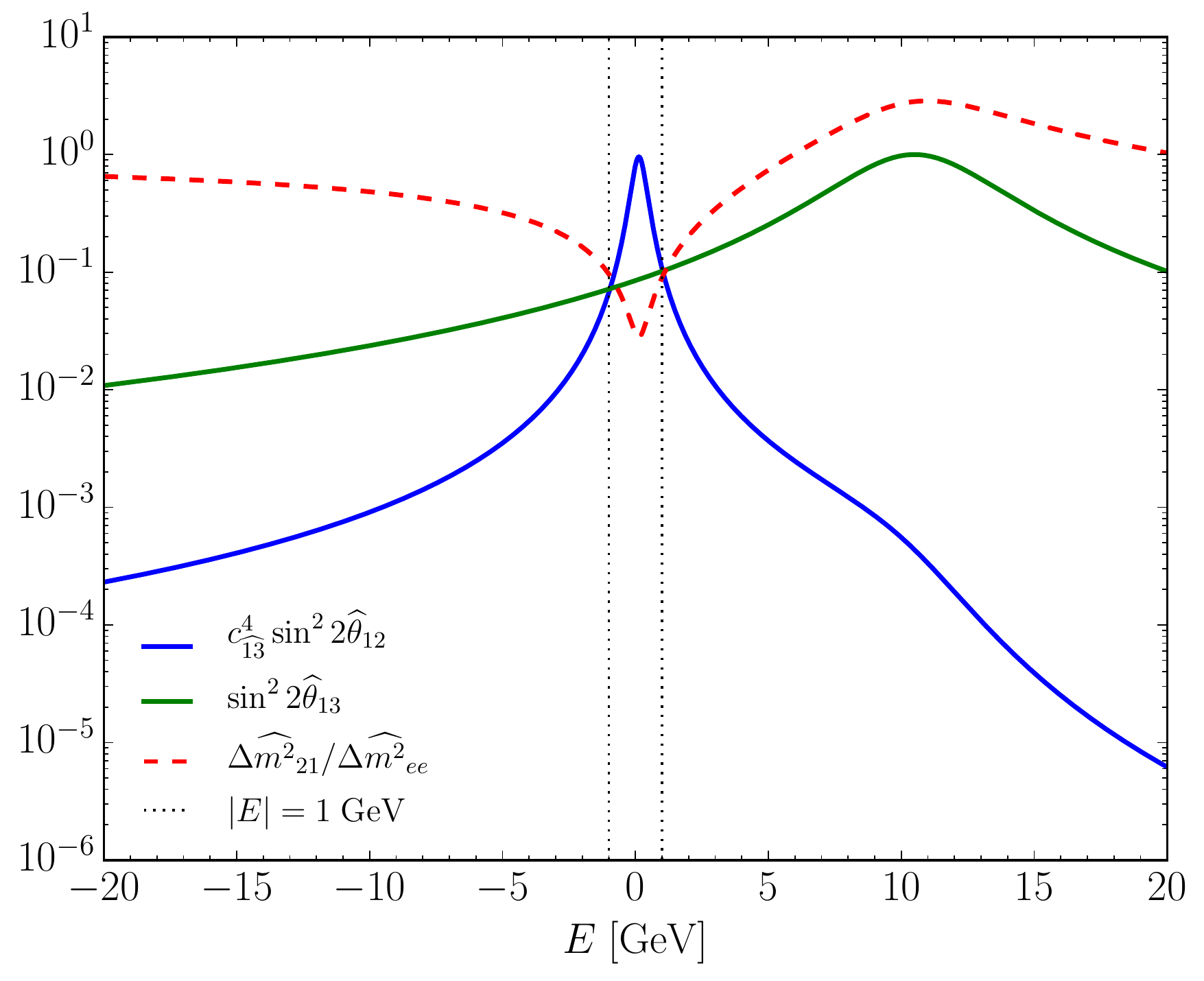}
\caption{An approximation of the size of the 21 term (numerator in blue, denominator in green) away from vacuum.
For $|E|\gtrsim5$ GeV we see that $R_2\ll1$.
See eq.~\ref{eq:R2} in the text.
We also show the ratio of the mass squared differences in matter in red.}
\label{fig:R2}
\end{figure}

First, if $\Delta\wh{m^2}_{21}\ll|\Delta\wh{m^2}_{ee}|$ then at the $n^{\text{th}}$ oscillation minimum the ratio $R_1$ of the 21 term to the $ee$ term is 
\begin{align*}
R_1&=\frac{C(E)c_{\wh{13}}^4\sin^22\wh\theta_{12}\sin^2\wh\Delta_{21}}{\sin^22\wh\theta_{13}}\\
&\approx\frac{\Delta m^2_{21}}{\dmsqee}[(2n-1)\pi/2]^2\left[C(E)c_{\wh{13}}^4c_{(\wh\theta_{13}-\theta_{13})}^2\frac{\sin^2\Delta_{21}}{\Delta_{21}^2}\right]\,,
\end{align*}
where the approximation uses the DMP zeorth order expression, the $\wh\theta_{13}\approx\wt\theta_{13}$ approximation of eq.~\ref{eq:s213m}, and $s_{13}^2\approx\Delta m^2_{21}/\dmsqee$.
The $C(E)$ term contains the effect of combining the $\wh\Delta_{31}$ and $\wh\Delta_{32}$ terms and is just under one within a few GeV of the vacuum.
Since all of the terms in the right square bracket are $<1$,
\begin{equation}
R_1\approx\frac{\Delta m^2_{21}}{\dmsqee}[(2n-1)\pi/4]^2\,.
\end{equation}
We numerically confirmed that eq.~\ref{eq:R} is correct to within $\sim10\%$ near vacuum as shown in fig.~\ref{fig:Error_21}.

The second case is when $\Delta\wh{m^2}_{21}\simeq|\Delta\wh{m^2}_{ee}|$, which occurs away from vacuum.
In this case we compare the ratio $R_2$ of the coefficients which is
\begin{equation}
R_2=\frac{c_{\wh{13}}^4\sin^22\wh\theta_{12}}{\sin^22\wh\theta_{13}}=\frac{|\wh{U}_{e1}|^2|\wh{U}_{e2}|^2}{|\wh{U}_{e3}|^2(1-|\wh{U}_{e3}|^2)}\,.
\end{equation}
Away from vacuum, $\wh\theta_{12}\simeq\pi/2$ ($0$) for neutrinos (anti-neutrinos) (see e.g.~fig.~1 of \cite{Denton:2016wmg}) which makes the numerator of $R_2$ very small.
The remaining part is $1/(4\tan^2\wh\theta_{13})$.
This part is large only when $\wh\theta_{13}\to0$.
Since $\wh\theta_{12}\to0$ faster than $\wh\theta_{13}$, we always have $R_2\ll1$ as desired.
See fig.~\ref{fig:R2} for a numerical verification that $R_2$ is small away from the vacuum.

\bibliography{dmsqee_matter}

\end{document}